\documentclass[prd,preprint,amsmath,nofootinbib,superscriptaddress]{revtex4}
\usepackage{graphicx}
\usepackage{bm}
\usepackage{epsfig}
\usepackage{color}
\usepackage{colordvi}

\newcommand{\beq}{\begin{equation}}
\newcommand{\eeq}{\end{equation}}
\newcommand{\bea}{\begin{eqnarray}}
\newcommand{\eea}{\end{eqnarray}}
\begin{document}

%%%%%%%%%%%%%%%%%%%%%%%%%%%%%%%%%%%%%%%%%%
%Define Title, Author, Address, Preprint#

\title{\boldmath 
Dissipative Effects in the Worldline Approach to Black Hole Dynamics\\
%\ \mydate 
}

\author{Walter D. Goldberger}
\affiliation{Department of Physics, Yale University, New Haven, CT 06520}
\author{Ira Z.~Rothstein}
\affiliation{Department of Physics, Carnegie Mellon University,
    Pittsburgh, PA 15213  \vspace{0.3cm}}
%%%%%%%%%%%%%%%%%%%%%%%%%%%%%%%%%%%%%%%%%%
\begin{abstract}

We derive a long wavelength effective point particle description of four-dimensional Schwarzschild black holes.   In this effective theory, absorptive effects are incorporated by introducing degrees of freedom localized on the worldline that mimic the interaction between the horizon and bulk fields.  The correlation functions of composite operators in this worldline theory can be obtained by standard matching calculations.  For example, we obtain the low frequency  two-point function of  multipole worldline operators by relating them to the long wavelength graviton black hole absorptive cross section.   The effective  theory is then used to predict  the leading effects of absorption in several astrophysically motivated examples, including the dynamics of non-relativistic black hole binary inspirals  and the motion of a small black hole in an arbitrary background geometry.   Our results can be written compactly in terms of absorption cross sections, and can be easily applied to  the dissipative dynamics of any compact object, e.g. neutron stars.  The relation of our methodology to that developed in the context of  the AdS/CFT correspondence is discussed.

 \end{abstract}

\maketitle

Understanding the dynamics of black holes in four dimensions has become a topic  of phenomenological interest.  The next generation of gravitational wave detectors will be sufficiently sensitive to probe effects arising from the dynamics of the horizon.   In particular, horizon dynamics in black hole binary star systems, including tidal deformation and absorption,  will leave an imprint on the measured signals.   For instance, it has been estimated in refs.~\cite{pheno} that the effects of horizon dissipation in the large mass black hole binary inspiral events seen by LISA can be large enough to account for about $5\%$ of the duration of the inspiral phase.   Reliable templates of the gravitational radiation waveforms emitted by these systems, which are necessary for interpretation of the LISA data, require that theoretical calculations be carried out to very high accuracy, making such effects relevant.

The problem of calculating the merger of two black holes as their orbits decay due to gravitational radiation is in general a difficult strong coupling problem, tractable only by numerical simulations.   However, there are several limits of astrophysical interest in which a small expansion parameter arises, and which can therefore be treated by analytical methods.  These include, for instance, the non-relativistic (or ``post-Newtonian") limit, and the extreme mass ratio limit, in which one of the black holes is much lighter and therefore moves approximately on a geodesic of the geometry generated by its heavier companion.       

In both of these situations, the characteristic gravitational fields in which at least one of the black holes move are typically long wavelength compared to the horizon size.   It is therefore convenient to treat the small black holes as point particles, and to describe their motion in terms of the worldline of some suitable center of energy coordinate.   However, the limit of singular delta function sources is problematic for conventional methods of solving the Einstein equations for binary systems, since point particle sources lead to singularities in the curvature invariants of the many-body spacetime which are difficult to handle by the tools of differential geometry.

 In a previous paper~\cite{GnR}, we constructed an effective field theory (EFT) method for systematically calculating gravitational wave observables within a point particle worldline approximation.  In this approach, the black holes (or any other compact object) are described by generalized worldline actions that include all possible terms consistent with the general coordinate invariance of general relativity.   By including all such operators it is possible to  (a) consistently renormalize all short distance divergences that may arise in calculations, and (b) systematically account for finite size (i.e., tidal) effects.   Furthermore, the coefficients of these non-minimal worldline operators can be obtained by matching calculations in which one compares the interactions of gravitons with an isolated black hole (for instance graviton scattering amplitudes) to those calculated in the point particle theory.
 
The effective point particle action of ref.~\cite{GnR} is adequate for obtaining a point particle description of non-dissipative tidal effects.   However, it cannot describe dissipative effects that may arise in realistic astrophysical sources, such as the absorption of orbital gravitational energy by the horizons of black holes within a binary star.   In this paper, we extend the EFT formalism to include dissipation.    
 
The presence of dissipation requires on general grounds the existence of a large number of gapless degrees of freedom~\cite{fd} and therefore cannot be described by a worldline action with just particle velocity and spin degrees of freedom\footnote{In fact, one may attempt to reproduce dissipation by including local operators with complex coefficients (leading to non-unitary time evolution and therefore dissipation).  However, it is possible to show that at least for Schwarzschild black holes, it is impossible to find non-Hermitian worldline operators that can reproduce the known energy dependence of absorption amplitudes for real valued fields with spin-$0,1$ or with spin-2.}.   It becomes necessary to introduce an additional number of worldline localized degrees of freedom that couple to gravity (as well as other four-dimensional fields) and can therefore account for the energy loss across the black hole horizon within a point particle description.   

While clearly these new worldline modes arise from a more complete, perhaps quantum mechanical or finite temperature,  description of the black hole (for instance, as the dimensional reduction of the degrees of freedom living on the stretched horizon~\cite{kip,susskind}), such a formulation is not necessary for our purposes here.  Rather, we will simply assume that these localized degrees of freedom couple to bulk fields through composite operators, labeled by the representation of the invariance group $SO(3)$ of symmetries in the rest frame of the particle, in a way that respects general covariance as well as worldline reparametrization invariance.  Thus these degrees of freedom can be interpreted in the bulk theory as dynamical multipole moments~\cite{geroch} in the local asymptotic rest frame of the particle.  Model independence is achieved by including all couplings and operators consistent with these symmetries.  

It is possible to calculate the relevant gravitational wave observables completely in terms of the $n$-point correlation functions of the multipole operators, which can be extracted by a standard matching calculation to observables of the Schwarzschild solution.   As we show explicitly in this paper, to leading order in the size of the black holes, the relevant quantities are the two-point function of the multipoles.  These can be obtained by calculating the amplitude for absorption of low energy modes of bulk, four-dimensional fields by the particle worldline, and comparing it to the analogous result obtained in the full black hole background~\cite{starobinski,page}.

As an example of these methods, we calculate the rate of horizon absorption of gravitational energy in black hole binary systems.   We first consider a toy model with an electromagnetically charged point particle gravitationally bound to black hole, taking the test charge to be for simplicity in a non-relativistic orbit.   In this scenario, we calculate the amount of electromagnetic energy absorbed by the black hole horizon.  In this case, the mass gained by the black hole can be qualitatively attributed to the coupling between the field of the test particle and a dynamically induced electric dipole moment on the black hole.   To our knowledge this result has not appeared elsewhere in the literature.  We then turn to the gravitational case.  In particular, we calculate the rate of mass change for a small black hole moving in an arbitrary background gravitational field, reproducing the recent result of ref.~\cite{poisson1}. We also calculate the leading contribution to mass increase for a pair of comparable mass black holes in the post-Newtonian approximation~\cite{alvi,poisson2}.  Since our results depend only on the graviton absorption cross section of the binary constituents, they can be easily applied generalized to other astrophysically relevant systems, for instance neutron star binaries.   The general formula for absorption in a non-relativistic binary star (neglecting contributions due to spin) is given in Eq.~(\ref{eq:general}).  The more phenomenologically relevant case of spinning compact binaries will be taken up elsewhere. 

On a more formal note, one may notice that the methodology discussed above is reminiscent of ideas used in the context of the AdS/CFT correspondence.   Some early evidence for this type of duality was given by Klebanov~\cite{Klebanov}, who pointed out that the low energy limit of classical scalar absorption by black 3-brane solutions of IIB supergravity is reproduced by an analogous \emph{quantum mechanical} process in which the absorption is attributed to the excitation of modes in the worldvolume theory of a D3 brane in flat spacetime by the incoming bulk field.  These results were later extended to include supergravity black holes in four and five dimensions with AdS near horizon geometries~\cite{teo}.   These examples are analogous to the situation here,  where we also calculate gravitational dissipative effects by rewriting them in terms of correlation functions in a non-gravitational localized theory.  However, unlike these string theoretic scenarios,  where the worldvolume theory is known (for the calculation of ref.~\cite{Klebanov} it is ${\cal N}=4$ super-Yang-Mills theory with gauge group $U(N)$), here we are only able to determine the correlation functions of the dual worldline theory by matching onto processes in the full black hole background.  Nevertheless, we find it quite interesting that, despite the geometrical differences between four-dimensional Schwarzschild black holes and string theory black holes, the basic methodology applies in an almost identical fashion. In this paper, however, the existence of a dual field theory follows simply from the basic tenets of effective field theory. This  ``universality'' of dual theory approaches is reviewed in \cite{carliprev}.

  \section {Black hole electrodynamics}

To begin with we consider the electrodynamics of black holes, in particular the absorption of photons by the black hole horizon, in the limit where all external fields have wavelength larger than the black hole radius $r_s$.   Following the ideas developed in~\cite{GnR}, in this limit the non-dissipative interactions of the black hole with photons can be systematically described in terms of an effective point particle action constructed from the relevant degrees of freedom in the problem:  the black hole worldline $x^\mu(\tau)$ and the gauge field $A_\mu(x)$.   By including in the action all operators consistent with the symmetries (gravitational and electromagnetic gauge invariance, worldline reparametrizations) we can systematically account for all finite size effects associated with black hole electrodynamics order by order in the parameter $r_s\omega\ll 1$, where $\omega$ denotes the typical frequency of the external fields.    The first few terms in this derivative expansion are given by 
  \begin{equation}
  \label{eq:localEM}
S_{eff} = e Q\int d x^\mu A_\mu + {\alpha\over 2}\int d\tau E^\mu E_\mu + {\beta\over 2}\int d\tau B^\mu B_\mu +\cdots,
 \end{equation}
  where $E_\mu=v^\nu F_{\nu\mu}$, $B_\mu={1\over 2}\epsilon_{\mu\alpha\beta\sigma} v^\sigma F^{\alpha\beta}$ and $v^\mu=dx^\mu/d\tau$.  We will only consider neutral black holes, $Q=0$, in what follows.  By dimensional analysis, we expect $\alpha,\beta$ to be proportional to a typical length scale $r_s$, and the coefficients of operators constructed with more derivatives acting on $F_{\mu\nu}$ (not shown) scale like more powers of $r_s$, leading to corrections to observables that are suppressed by powers of $r_s\omega$ relative to those shown here.

While this effective action can be used to reproduce elastic photon scattering amplitudes  to arbitrary order in $r_s\omega$, it cannot be used to reproduce black hole absorption.  On general grounds, dissipation naturally implies the existence of a large number of modes\footnote{Here, ``large" means large enough that the Poincare recurrence time is long compared to the other scales in the problem.} with energy near that of the vacuum.  As such, these degrees of freedom must be included in the effective theory since they may lead to non-local effects that cannot be reproduced by Eq.~(\ref{eq:localEM}) alone.

In general we do not know the explicit dynamics of these additional degrees of freedom.  However, we may use the symmetries of the black hole solution to determine the spectrum of possible composite operators in this theory.  In the case of a  Schwarzschild black hole interacting with a field theory that conserves parity, the operator spectrum can be classified in terms of representations of the isometry group $SO(3)$ of the background as well as a parity eigenvalue.  Thus the Lagrangian that describes the interaction of the black hole with the electromagnetic field is of the form
  \begin{equation}
  \label{eq:BHQED}
S_{int}=e Q\int d x^\mu A_\mu -\int d\tau p_a(\tau) E^a(\tau) - \int d\tau m_a(\tau) B^a(\tau) + \cdots,
  \end{equation}     
  where $a=1,2,3$ are $SO(3)$ indices, and  $p_a(\tau), m_a(\tau)$ are composite operators corresponding to electric and magnetic parity electromagnetic dipole moments.   In this equation, $E^a=e^a_\mu E^\mu$, $B^a=e^a_\mu B^\mu$ are the electric and magnetic fields measured in the frame of the black hole, and the set of vectors $e^\mu_a$ constitute a frame orthogonal to the velocity $v^\mu$ and satisfying $g_{\mu\nu} e^\mu_a e^\nu_b = -\delta_{ab}$, $\delta_{ab} e^a_\mu e^b_\nu = -g_{\mu\nu}+v_\mu v_\nu$.  Note that the isometry group of the gravity background acts in the worldline theory as a global symmetry.   This relation is analogous to what occurs in the AdS/CFT correspondence, where the isometries of the AdS background map onto the global symmetries of the CFT.

\begin{figure}[!t]
\label{subD}
%\centerline{\scalebox{0.99}{\includegraphics*[60, 290][200,830]{FP.eps}}}
\centerline{\scalebox{0.99}{\includegraphics{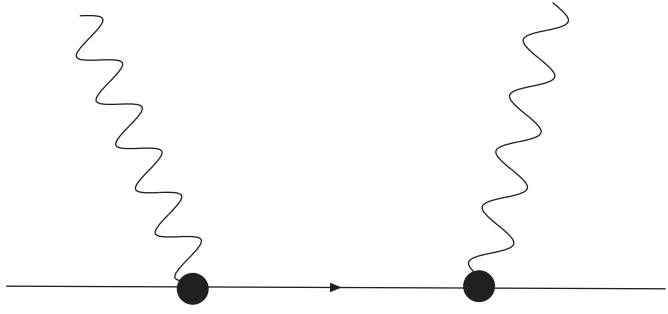}}}
\vskip-0.3cm
\caption{Feynman diagram whose imaginary part gives the leading order contribution to the absorptive cross section. The dots correspond to insertions of leading multipole worldline operators.}
\label{fig1} 
\end{figure}

Given the correlation functions of worldline operators, one may determine any observable involving black holes.  Although we do not know how to compute these Green's functions from first principles, it is possible to relate them to known black hole observables.  The idea is to match scattering amplitudes calculated using Eq.~(\ref{eq:BHQED}) to known results calculated in a Schwarzschild background.   For instance, to determine the effects of horizon dissipation it is sufficient to compare the known low energy absorption cross section for polarized photons, calculated by solving Maxwell's equations around a black hole~\cite{starobinski,page},
\begin{equation}
\label{eq:pageQED}
\sigma_{abs,p}(\omega)= {4\pi\over 3} r^4_s\omega^2+{\cal O}(\omega^3),
\end{equation} 
(where $p=\pm 1$ labels the polarization of the incoming state) to the predictions of Eq.~(\ref{eq:BHQED}).  In terms of the two-point functions $\langle p_a p_b  \rangle$, $\langle m_a m_b\rangle$, Eq.~(\ref{eq:BHQED}) predicts, in the black hole rest frame (see Fig.~\ref{fig1})
\begin{equation}
\label{eq:qedabs}
\sigma_{abs}(\omega)={1\over 2\omega}\int dx^0 e^{-i \omega x^0} \left[\omega^2 \epsilon^*_a \epsilon_b\langle p_a (x^0)p_b(0))  \rangle  + ({\bf k}\times {\bf\epsilon}^*)_a ({\bf k}\times {\bf\epsilon})_b \langle  m_a (x^0)m_b(0)\rangle\right],
\end{equation}
where $\epsilon_a$ is the incoming photon polarization, $\bf k$ is the photon momentum 3-vector,  and the expectation values are taken in the ground state $|\Omega\rangle$ of the worldline theory.  

It is actually more convenient for us to work in terms of the time-ordered correlators of dipole operators, rather than the expectation values appearing in Eq.~(\ref{eq:qedabs}).   Using the relation
\begin{equation}
\int dx^0 e^{-i \omega x^0} \langle  O_a(0) O_b(x^0)\rangle = 2\mbox{Im} i \int dx^0 e^{-i \omega x^0} \langle T O_a(0) O_b(x^0)\rangle,
\end{equation}   
valid for $\omega>0$ ($O_a$ is either $p_a$ or $m_a$), as well as rotational invariance
\begin{equation}
\label{eq:EMTOP}
\int dx^0 e^{-i \omega x^0} \langle T p_a(0) p_b(x^0)\rangle =  -i \delta_{ab} F(\omega),
\end{equation}
we find
\begin{equation}
\sigma_{abs,p}(\omega)= 2\omega\mbox{Im}  F(\omega)
\end{equation}
for each polarization state of the photon. This equation is a special case of the fluctuation-dissipation theorem~\cite{fd}. Note that we have used the equality of the time ordered two-point functions of $p_a$ and $m_a$, which follows from the duality invariance $E^a\rightarrow - B^a$, $B^a\rightarrow E^a$ of the Maxwell equations in the Schwarzschild vacuum.  For the effective theory of Eq.~(\ref{eq:BHQED}) to reproduce the duality invariance of the full theory, we must have $p_a\rightarrow -m_a$ under a duality flip, guaranteeing the equality of magnetic and electric correlators.  Comparing to Eq.~(\ref{eq:pageQED}), this yields
\begin{equation}
\label{eq:im2pt}
\mbox{Im}  F(\omega) = {2\pi\over 3} r^4_s   |\omega|,
\end{equation}
where we have used the fact that $F(-\omega)=F(\omega)$, which follows from the definition, Eq.~(\ref{eq:EMTOP}).  The non-absorptive part of the two-point function can be obtained in an analogous manner, by computing the amplitude for photon elastic scattering off the black hole. By dispersion relation arguments, this term is even in $\omega$ and therefores lead to a term in the amplitude that can be reproduced by one of the terms in Eq.~(\ref{eq:localEM}).

The utility of knowing correlators such as $\langle p_a p_b\rangle$ is that they are universal.  Having extracted them from a simple process such as the one we just considered, they can be applied in more complex (and realistic) dynamical situations involving multiple black holes interacting with electromagnetic and gravitational fields.   All calculations can be done in the point particle approximation, where the dynamics simplifies, and finite size effects are systematically incorporated by higher-dimensional worldline operators in Eq.~(\ref{eq:localEM}) and Eq.~(\ref{eq:BHQED}).  Although in general one may encounter divergences in the point particle limit, these can be renormalized into the coefficients of the operators in Eq.~(\ref{eq:localEM}).   This is explained in detail in ref.~\cite{GnR}.

Physically, the relevant observable in the gravitational many body problem is the flux of energy radiated to infinity in massless fields (photons and gravitons).   Following~\cite{GnR}, this observable can be calculated from the quantity $S_{eff}[x_a]$, defined by the expression
\begin{equation}
\label{eq:PI}
\exp[iS_{eff}[x_a]] = \int {\cal D} h_{\mu\nu} {\cal D} A_{\mu} {\cal D} X_a\exp[iS[x_a,X_a,A,h]],
\end{equation}
In this equation, $S[x_a,X_a,A,h]$ is the action that describes the interaction of photons and gravitons with the degrees of freedom $X_a$ on the worldline $x_a$ of the $a$-th black hole.    Besides the usual gravitational term $-2 m^2_{Pl} \int d^4 x \sqrt{g} R,$ $S[x_a,X_a,A,h]$ contains the (unknown) worldline Lagrangian for the $X$ terms, as well as the action for the point particles, including the finite size terms of Eq.~(\ref{eq:localEM}) and Eq.~(\ref{eq:BHQED}).  

$S_{eff}[x_a]$ has a graphical interpretation as the sum of connected diagrams with no external graviton (and photon) lines\footnote{As usual, diagrams with loops of gravitons are quantum effects suppressed by powers of the typical energy scale over the Planck mass.}.   The real part of $S_{eff}[x_a]$ gives rise to the conservative terms in the multi-body Lagrangian, from which the equations of motion can be extracted.  The imaginary part accounts for the fact that energy in the electromagnetic and gravitational fields is transferred either to infinity or to the worldline degrees of freedom $X_a$.  In other words, the system of massless fields is open and therefore its time evolution is non-unitary.  In practice, one may distinguish between the energy lost to infinity and the energy absorbed by the degrees of freedom $X_a$ by examining the individual diagrams contributing to $\mbox{Im} S_{eff}[x_a]$.   Specifically,
\begin{equation}
\label{eq:ugtp}
{2\over T} \mbox{Im} S_{eff}[x_a]  =  \int dE {d\Gamma\over dE},
\end{equation}
gives the rate of energy loss either to infinity or to the black holes.  One can obtain the total power loss as $dP= E d\Gamma.$

As an application, consider a charged point particle in orbit around a black hole (orbital radius much larger than the Schwarzschild radius).   Given Eq.~(\ref{eq:im2pt}) it is easy to calculate how much of the radiation emitted by the accelerated point particle is absorbed by the black hole.  In this case the dynamics is given, in the black hole rest frame where the frame $e^a_\mu$ lines up with the global flat space coordinates, by 
\begin{eqnarray}
\nonumber
S &=&-m\int d\tau\sqrt{1+h_{\mu\nu} v^\mu v^\nu} + e\int dx^\mu A_\mu(x)\\
& &  - M\int dt\sqrt{1+h_{00}} + S[X] -\int dt p_a(t) E^a(t)-\int dt m_a(t) B^a(t).
\end{eqnarray}
\begin{figure}[!t]
%\centerline{\scalebox{0.99}{\includegraphics[60, 390][200,630]{FP.eps}}}
\centerline{\scalebox{0.99}{\includegraphics{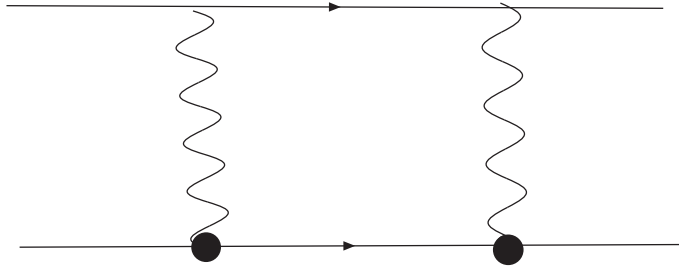}}}
\vskip-0.3cm
\caption{Leading order contribution to the absorptive potential. The dots correspond to insertions of the leading worldline multipole operators.}
\label{fig2} 
\end{figure}
The leading order term in the effective action is simply,
\begin{equation}
\label{eq:LO}
i S_{eff}[x] = -{m M\over 4 m^2_{Pl}} \int d\tau dt v^\mu(\tau) v^\nu(\tau)\langle T h_{\mu\nu}(x(\tau)) h_{00}(0,t)\rangle +\cdots,
\end{equation}
while the term of  Fig.~\ref{fig2} captures the dissipative features of the black hole,
\begin{equation}
\label{eq:QEDbox}
iS_{eff}[x]= {e^2\over 2}\int dx^\mu d{\bar x}^\nu dt d{\bar t}\langle T p_a(t) p_b({\bar t)}\rangle \left[\langle T A_\mu(x) E^a(t)\rangle \langle T A_\nu({\bar x}) E^b({\bar t})\rangle+(E\leftrightarrow B)\right]+\cdots.
\end{equation}
For simplicity we will take the point particle to be in a non-relativistic orbit.   In this case, to leading order in the three-velocity $v\ll 1$,  Eq.~(\ref{eq:LO}) reduces to the Lagrangian for a NR particle in a Newtonian potential $V(r)=-G_N m M/r$.   In the NR limit, only the electric contribution to Eq.~(\ref{eq:QEDbox}) survives (the magnetic field being suppressed by an additional power of $v$).   Using
\begin{equation}
\langle T A_0(x^0,{\bf x}) E_i(0)\rangle = {1\over 4\pi}\delta(x^0) {{\bf x}_i\over |{\bf x}|^3},
\end{equation}
Eq.~(\ref{eq:QEDbox}) becomes
\begin{equation}
iS_{eff}[x]={e^2\over 32\pi^2}\int dx^0 d{\bar x}^0 \langle T p_i(x^0) p_j({\bar x}^0)\rangle d_i(x^0) d_j({\bar x^0})+\cdots,
\end{equation}
where we have defined ${\bf d}(t)={\bf x}(t)/|{\bf x}(t)|^3$.  This gives
\begin{equation}
S_{eff}[x]=-{e^2\over 32\pi^2}\int{d\omega\over 2\pi} |d_i(\omega)|^2 F(\omega)+\cdots,
\end{equation}
and by Eq.~(\ref{eq:ugtp}), 
\begin{equation}
{dP_{abs} \over d\omega}=-{1\over T} {e^2\over 24\pi^2} r^4_s \omega^2 |d_i(\omega)|^2.
\end{equation}
Integrating over the physical region $\omega>0$ we find
\begin{equation}
P_{abs}=-{e^2 \over 24\pi}r^4_s\langle\dot{\bf d}\cdot\dot{\bf d}\rangle,
\end{equation}
where the brackets denote a time average.  Comparing to the leading order dipole radiation formula, this shows that the absorption of electromagnetic radiation by the black hole is an effect that is down by $v^6$ in the NR limit.

Although here we have only considered the leading order NR expression for electromagnetic absorption, it is possible to systematize the expansion to all orders in $v$, by a trivial modification of the velocity power counting rules developed in ref.~\cite{GnR} for gravity.   In particular, to obtain manifest velocity scaling in the electromagnetic sector~\cite{TASI,ben}, one would have to decompose the photon $A_\mu$ into a potential term with energy-momentum $(v/r,1/r)$ and a radiation piece with momentum components $(v/r,v/r)$.  In addition, we would have to assign the power counting rule $p_a\sim \sqrt{L} v^{7/2}/m_{Pl}$ in the non-relativistic limit ($L=mvr$ is the typical angular momentum of the system) to keep track of powers of $v$ arising from higher order insertions of the dipole operator.   One would also have to include insertions of higher multipole operators.

Finally, we stress that the analysis above is completely general.   It would apply, for instance, in the analysis of a finite extent condensed matter system interacting with long wavelength electromagnetic fields.   In that case, the two-point functions $\langle p_a p_b\rangle$, $\langle m_a m_b\rangle$ would describe to the electric and magnetic susceptibilities of the material.

\section{Gravitational Dissipation}

The absorption of gravitational energy by black holes can be understood in a manner analogous to our discussion of electrodynamics.   The non-dissipative part of the response of a black hole to external gravitational fields is encoded by a local effective point particle action of the form
\begin{equation}
S_{eff}[x]=-m\int d\tau + {\alpha\over 2}\int d\tau E^{\mu\nu} E_{\mu\nu}+{\beta\over 2}\int d\tau B^{\mu\nu}  B_{\mu\nu} +\cdots.
\end{equation}
In this equation we assume for simplicity that the background field is a vacuum solution, so that all operators can be written in terms of the ten independent components of the Weyl tensor:  the five electric-type parity components $E_{\mu\nu}=C_{\mu\alpha\nu\beta} v^\alpha v^\beta$, and the five magnetic components $B_{\mu\nu}={1\over 2}\epsilon_{\mu\alpha\beta\rho} {C^{\alpha\beta}}_{\nu\sigma} v^\rho v^\sigma$.  By dimensional analysis, we expect $\alpha,\beta$ to depend cubically on the internal length scale of the system.  Other operators arising at this order, involving the Ricci curvature, can be removed by a field redefinition of the metric.  

As in the previous section, we take the presence of absorptive processes to signal the existence of a large number of worldline modes that interact with gravitons.  Although we do not know what this worldline theory is, in the spirit of effective field theory we determine the interactions by writing down all possible operators consistent with the symmetries.   The simplest possible couplings, involving two derivatives, are given by
\begin{equation}
\label{eq:BHgrav}
S=-\int d\tau Q^E_{ab} E^{ab}-\int d\tau Q^B_{ab} B^{ab}+\cdots,
\end{equation}
where $E^{ab}=e^a_\mu e^b_\nu E^{\mu\nu}$, $B^{ab}=e^a_\mu e^b_\nu B^{\mu\nu}$.  The operators $Q^{E,B}_{ab}$ are electric and magnetic quadrupole type parity operators composed of the worldline degrees of freedom.    Terms with more derivatives, which are suppressed in the low energy limit, are not shown.  Using this equation, the graviton absorption cross section is given by
\begin{equation}
\sigma_{abs}(\omega)={\omega\over 8 m^2_{Pl}}\int dx^0 e^{-i\omega x^0}\left[
\omega^2 \epsilon^*_{ab}\epsilon_{cd} \langle Q^E_{ab}(0) Q^E_{cd}(x^0) \rangle + ({\bf k}\times \epsilon^*)_{ab}  ({\bf k}\times \epsilon)_{cd} \langle Q^B_{ab}(0) Q^B_{cd}(x^0)\rangle\right],
\end{equation}
where $({\bf k}\times\epsilon)_{ab}={{\bf \epsilon}_{acd}} {\bf k}_c\epsilon_{db}$.  In terms of the two-point function
\begin{equation}
\label{eq:g2pt}
\int dx^0 e^{-i\omega x^0} \langle T Q^E_{ab}(0) Q^E_{cd}(x^0)\rangle = -{i\over 2}\left[\delta_{ac}\delta_{bd} +\delta_{ad}\delta_{bc} -{2\over 3}\delta_{ab}\delta_{cd}\right] F(\omega),
\end{equation}
the cross section reads
\begin{equation}
\sigma_{abs,p}(\omega)={\omega^3\over 2 m^2_{Pl}}\mbox{Im}  F(\omega).
\end{equation}
Matching to the result in ref.~\cite{starobinski,page},
\begin{equation}
\label{eq:gcross}
\sigma_{abs,p}(\omega)={1\over 45} 4\pi r^6_s\omega^4,
\end{equation}
we find $\mbox{Im}  F(\omega) = 16 G^5_N m^6 |\omega|/45$.   Here we have used the equality of the magnetic and electric correlators, which follows from the duality invariance $E_{ab}\rightarrow -B_{ab}$, $B_{ab}\rightarrow E_{ab}$ of the Teukolsky equation.

As a simple application of Eq.~(\ref{eq:BHgrav}), we consider the motion of a black hole in a background vacuum spacetime, taking the curvature length scale ${\cal R}$ much larger than the black hole radius $r_s$.   In order to determine how much of the gravitational energy of the background is absorbed by the black hole (as measured in the frame of the hole), we compute the functional $S_{eff}[x]$ defined by Eq.~(\ref{eq:PI}).   Ignoring gravitational fluctuations, we can simply substitute the background values of $E_{ab}, B_{ab}$ into Eq.~(\ref{eq:BHgrav}).  In this case the leading term in the path integral that generates $S_{eff}[x]$ arises from two insertions of the operators $Q^{E,B}_{ab}$,
\begin{equation}
S_{eff}[x]=-m\int d\tau +{i\over 2}\int d\tau d{\bar \tau}\langle T Q^E_{ab}(\tau) Q^E_{cd}({\bar \tau})\rangle\left[E^{ab}(\tau) E^{cd}({\bar\tau})+B^{ab}(\tau) B^{cd}({\bar\tau})\right]+\cdots
\end{equation}
From the previous section, we expect the imaginary part of this equation to be related to the power absorbed by the black hole. In particular this gain in energy by the black hole is the measured change in the mass by a ``local'' observer~\cite{hartle,alvi}.  The increase in mass by the black hole will affect its motion through the background spacetime, inducing a correction in the phase of the radiated gravitational waveform measured by an observer at infinity~\cite{pheno}.  Calculating in the black hole frame we find,
\begin{equation}
\mbox{Im} S_{eff}[x]={16\over 45} G^5_N m^6\int^\infty_0{d\omega\over 2\pi} \, \omega \left[E_{ab}(\omega) E^{ab}(-\omega) + B_{ab}(\omega) B^{ab}(-\omega)\right],
\end{equation}
and therefore
\begin{equation}
\label{eq:bpower}
P_{abs} = {16\over 45} G^5_N m^6 \langle {\dot E}^{ab}{\dot E}_{ab} +  {\dot B}^{ab}{\dot B}_{ab}\rangle,
\end{equation}
were ${\dot E}_{ab}=dE_{ab}/d\tau$. Ignoring spin effects, the three-frame $e^a_\mu$ is parallel transported along the worldline so that ${\dot E}^{ab}= e^a_\alpha e^b_\beta (v\cdot D)  E^{\alpha\beta}$, and the above expression is covariant.   

It is also possible to use Eq.~(\ref{eq:BHgrav}) to account for the effects of absorption in the limit where two or more black holes are bound gravitationally in a non-relativistic orbit.   As a simple example, we consider the effects of absorption in a slowly inspiraling binary system of two black holes.  This can be done systematically, in the point particle limit, using the velocity power counting rules of ref.~\cite{GnR}.  From Eq.~(\ref{eq:g2pt})-(\ref{eq:gcross})  we see that $\langle Q^E_{ab} Q^E_{cd}\rangle\sim G^5_N m^6\omega^2$ and therefore, in the NR limit, with $\omega\sim v/r$,  $Q_{ab}^E\sim L v^4/m_{Pl}$, where $L=m v r$ is the typical angular momentum of the binary system.  In the nomenclature of ref.~\cite{GnR} we therefore have
\begin{equation}
\int d\tau Q^E_{ab} E^{ab}[H]\sim v^{13/2},
\end{equation}
from the leading term in the NR expansion of $E_{ab}$ in terms of potential graviton modes $H_{\mu\nu}$ (which have energy-momentum $\sim (v/r, 1/r)$ and thus generate instantaneous two-body interactions),
\begin{equation}
\label{eq:el}
E_{ij}={1\over 2}\partial_i\partial_j H_{00}+\cdots.
\end{equation}  
 Likewise, we find that the magnetic potential $B_{ab}$ is given to leading order by
 \begin{equation}
 \label{eq:mag}
B_{ij} = \epsilon_{irs}\left[\partial_r\partial_j H_{0s} + \partial_s\partial_j H_{0r} \right]+\cdots
 \end{equation}
 and therefore
\begin{equation}
\int d\tau Q^B_{ab} B^{ab}[H]\sim v^{13/2},
\end{equation}
in the NR limit as well.  Similar electric and magnetic terms linear in radiation modes $h_{\mu\nu}$ (which have on-shell momentum $(v/r,v/r)$ and are thus propagating degrees of freedom) are suppressed by an additional factor of $v^{5/2}$. 

By rotational invariance, $\langle Q^E_{ab}\rangle=0$ and the leading order absorptive contribution to $S_{eff}[x_a]$ is from a box diagram similar to the one we found in black hole electrodynamics, with two insertions of the coupling Eq.~(\ref{eq:BHgrav}) and two insertions of the leading order Newtonian interaction from the second particle.    Note that the magnetic coupling leading to graviton exchange between sources is suppressed by a factor of $v$ since $h_{0i}$ does not couple to leading order to $h_{00}$. Thus to get a non-vanishing magnetic contribution to the box diagram one needs two insertions of a vertex of the form $v_i h_{0i}$\footnote{The correlator with one electric and one magnetic quadrupole insertion vanishes in the worldline theory by parity invariance.}.

Given that the Newtonian interaction 
\begin{equation}
S=-\sum_a {m_a\over 2 m_{Pl}}\int dx^0 H_{00}+\cdots,
\end{equation}
scales as $\sqrt{L}$, the box diagram contributes to $\mbox{Im}S_{eff}$ a term that is of order $Lv^{13}$.  For comparison, the leading order diagram corresponding to radiation by the mass quadrupole of the binary system scales as $Lv^5$.   Thus absorption from NR black hole binaries is a $v^8$ effect in the case where the constituents are not spinning~\cite{poisson2}.   However, for spinning black holes this effect may be enhanced by up to three powers of $v$~\cite{Tagoshi}, leading to potentially observable effects in the LISA data~\cite{pheno}.   A calculation of this effect in the worldline theory, using the spin formalism developed in~\cite{porto}, is in progress.

Explicitly calculating the box diagram of Fig.~\ref{fig2}, we find that the absorptive term in $S_{eff}$ is
\begin{eqnarray}
\nonumber
iS_{eff}[x_1,x_2] &=& {m^2_2 \over 8 m^4_{Pl}} \int {dx^0_1}  {d{\bar x}^0_1} {dx^0_2}  {d{\bar x}^0_2}   \langle T H_{00}(x^0_2) E_{ij}(x^0_1)\rangle  \langle T H_{00}({\bar x}^0_2) E_{rs}({\bar x}^0_1)\rangle\\
& &{} \times \langle T Q^E_{ij}(x^0_1) Q^E_{rs}({\bar x}^0_2)\rangle+(1\leftrightarrow 2) + \cdots.
\end{eqnarray}
Using Eq.~(\ref{eq:el}) we find
\begin{equation} 
\langle T H_{00}(x^0,{\bf x}) E_{ij}(0) \rangle = {i\over 4\pi} \delta(x^0) \partial_i \partial_j {1\over |{\bf x}|},
\end{equation}
and  defining $q^{(a)}_{ij}(t)=\partial^a_i\partial^a_j |{\bf x}_{12}(t)|^{-1}$ $(a=1,2)$, we obtain
\begin{equation}
S_{eff}[x_1,x_2]= {1\over 4} G^2_N\sum_{a\neq b} \int {d\omega\over 2\pi} F_b(\omega)m^2_a |q^{(a)}_{ij}(\omega)|^2+\cdots.
\end{equation}
Thus the binding energy loss due to absorption is given by
\begin{eqnarray}
\nonumber
P_{abs} &=&  {16\over 45} G^7_N m^2_1 m^2_2\left\langle {1\over 2}\sum_{a\neq b} m^4_a {\dot q}^{(b)}_{ij} {\dot q}^{(b)}_{ij} \right\rangle\\
&=& {32\over 5} G_N^7m^6\mu^2\left\langle {{\bf v}^2\over |{\bf x}|^8} + 32 {({\bf x}\cdot{\bf v})^2\over |{\bf x}|^{10}}\right\rangle.
\end{eqnarray}
%due to the absorption of gravitational energy by the black holes.
  In the second line, we have used CM coordinates, $\sum_a m_a {\bf x}_a=0$, and defined ${\bf x}={\bf x}_1-{\bf x}_2$, $m=\sum_a m_a$, $\mu= m_1 m_2/m$.

Higher order relativistic corrections to this result follow directly by including velocity suppressed vertices from the point particle  and gravitational actions.   The systematic inclusion of such terms is standard, and is described in~\cite{GnR}.  There are also corrections from insertions of higher multipole worldline operators, although these are highly suppressed.  For instance, an operator of the form $\int d\tau Q^E_{abc} D^c E^{ab}$, which generates a correction to the graviton absorption cross section suppressed by $r^2_s\omega^2$ relative to Eq.~(\ref{eq:gcross}), scales as $v^{17/2}$ in the NR limit, giving a correction to the power absorption that is down by $v^{12}$ relative to the leading order quadrupole radiation.  Finally, there may be corrections from more than two insertions of the quadrupole operators.  Such terms involve the three-point and higher correlation functions of $Q^E_{ab}$.  However, given that the operator $\int d\tau Q^E_{ab} E^{ab}$ scales as $v^{13/2}$ such contributions are completely irrelevant.

\section{Speculations}

In the last two sections, we have introduced a formalism for including dissipation in the worldline description of a black hole by including a set of localized degrees of freedom whose internal dynamics account for the absorption of external field quanta.   Although we used the formalism merely as a bookkeeping device, it is interesting to assign a physical meaning  to these worldline modes as the dimensional reduction down to $0+1$ dimensions (i.e., the particle worldline) of a hypothetical two-dimensional theory localized on the stretched horizon~\cite{susskind}.  The possible existence of such a horizon theory has been described in many different contexts, mostly to assign a microphysical origin to black hole entropy~\cite{mathur}.
 
In the case of  AdS/CFT, the dual field theory is known in many instances.  In Klebanov's  work~\cite{Klebanov} the absorption of gravitons polarized parallel to a D3-brane in IIB string theory is reproduced by calculating the imaginary part of the two-point  function of the stress-energy tensor in the  D-brane worldvolume gauge theory, ${\cal N}=4$ supersymmetric Yang-Mills (SYM),
\beq
\sigma=\frac{\kappa^2}{\omega}\int d^4x e^{i\omega t}\langle [T_{xy}(x),T_{xy}(0)]\rangle.
\eeq
where $\kappa=\sqrt{8\pi G_{10}}$, with $G_{10}$ the ten dimensional Newton constant, and
the $xy$ plane is parallel to the D3-brane\footnote{The cross section is defined per unit longitudinal volume.}.  Perhaps the more relevant  analogy however, can be found in ref.~\cite{son} where the calculation of the two point function was done in the near-extremal limit
in which case the dual field theory is at finite temperature and relates the bulk absorptive cross section to the shear viscosity\footnote{Note that in this theory it is more difficult to  check the duality since supersymmetry no longer protects the correlator in the gauge theory.}.  A crucial aspect of these calculations is that the near-horizon geometry is a tensor product of an AdS space and a compact manifold.  However,  it has been shown that, even  without  supersymmetry~\cite{carliporig}, there exists a   dual field theory. This result follow from the fact that the asymptotic symmetry group for the BTZ black hole is a Virasoro algebra~\cite{brown} thus leading to a dual conformal field theory whose central
charge determines an entropy which agrees with the Bekenstein-Hawking result~\cite{strominger}.
It has has been suggested that this relation holds even for non-AdS geometries~\cite{carlipnoads,MS,sfestos}. Finally  we should note that there is related work on using quantum mechanics 
 to describe horizon degrees of freedom.

 In~\cite{kabat} the authors use the duality between the ten dimensional D0 brane  black hole and supersymmetric quantum mechanics~\cite{Itzhaki} to reproduce  the thermodynamic properties of the black hole by treating the strongly coupled quantum mechanical theory as a gas  of non-interacting quasi-particles.  
While we do not know the dual theory in the examples of this paper, we can extract the correlation functions of the theory in the low energy limit. The higher order terms in the full theory cross section will correspond to correlators of higher moment operators. In principle one could also extract the higher $n$-point correlators by considering scattering contributions from time ordered products including multiple insertions of operators, though it is not clear how one would disentangle these contributions from lower order insertions of higher multipoles.  Nonetheless it seems plausible that one could narrow down  classes of theories which could reproduce known correlators in an attempt to find the true horizon theory.

\section{Conclusions}

In this paper, we have presented a simple method for including the effects of absorption in the dynamics of compact objects interacting through long wavelength gravitational fields.  In this limit, one may use effective field theory techniques to  ``integrate out" the internal structure of the compact objects, treating them as point particles with generalized worldline actions whose terms encode their finite size.   While such terms are sufficient to give a simple yet systematic description of non-dissipative many body gravitational dynamics, they do not capture dissipative effects, such as the absorption of gravitational radiation by black hole horizons.  In order to include absorptive effects we have proposed introducing additional gapless degrees of freedom localized on the worldline which couple to external fields.

Using this method we determined the absorptive dynamics of black holes in various situations.   In particular we worked out the amount of electromagnetic energy absorbed by a charge in orbit around a neutral black hole.   We also calculated the amount of gravitational energy absorbed by a black hole in a  long wavelength background gravitational field, as well as the fraction of the gravitational radiation that gets absorbed rather than emitted to infinity by two black holes in a non-relativistic binary inspiral event, 
to leading order in the velocity $v$, finding agreement with calculations done by more traditional methods~\cite{alvi}.

Although we have only considered the theoretically clean case of black holes, our methods generalize easily to other objects of astrophysical interest.   For example for several objects in NR orbits, the power spectrum for absorption of gravitational energy over an observation time $T$ is given by
\begin{equation}
\label{eq:general}
{dP_{abs}\over d\omega} = -{1\over T} {G_N\over 64\pi^2} \sum_{a\neq b} {\sigma^{(b)}_{abs}(\omega)\over\omega^2} m^2_a |q^{(a)}_{ij}(\omega)|^2,
\end{equation}
where $\sigma^{(a)}_{abs}(\omega)$ is the graviton absorption cross section for each object in the system.  This formula can be used to predict the absorptive corrections to the gravitational wave flux seen by a detector, given a model for the internal structure in which $\sigma_{abs}(\omega)$ can be calculated.  

At higher orders one will encounter divergences due to the point particle approximation.
As was shown in \cite{GnR} these divergences will simply renormalize higher dimensional
interactions terms. These terms are also responsible for reproducing the finite size effects
of the objects. Thus there is no  impediment to calculating to any desired order using the
power counting rules developed in \cite{GnR}.

Finally, once the effects of spin are included in the effective theory~\cite{porto}, the dissipative dynamics of rotating black hole or neutron star systems can be worked out. These cases are more phenomenologically relevant than the examples discussed in this paper.

\centerline{\bf Acknowledgments}
The authors would like to thank D. Boyanovsky, T. Jacobson,  A. Manohar, E. Poisson,  and D. Son for helpful discussions.  WG is supported in part by the Department of Energy under Grant DE-FG02-92ER40704 and thanks the Aspen Center for Physics where some of this work was completed.   IZR is supported in part by the Department of Energy under Grants DOE-ER-40682-143 and DEAC02-6CH03000.

\end{document}